\documentclass[lettersize,journal]{IEEEtran}
\usepackage{amsmath,amsfonts}
\usepackage{algorithmic}
\usepackage{array}
\usepackage[caption=false,font=normalsize,labelfont=sf,textfont=sf]{subfig}
\usepackage{textcomp}
\usepackage{stfloats}
\usepackage{url}
\usepackage{verbatim}
\usepackage{graphicx}

\usepackage[active]{srcltx}
\usepackage{color}
\graphicspath{{figures/}}

\def\BibTeX{{\rm B\kern-.05em{\sc i\kern-.025em b}\kern-.08em
    T\kern-.1667em\lower.7ex\hbox{E}\kern-.125emX}}
\usepackage{balance}

\begin{document}

\title{On the degeneracy of whispering gallery modes in a high-Q sapphire microwave resonator}
\author{Vincent Giordano, Samuel Margueron
\thanks{V. Giordano, S. Margueron are with the Institute FEMTO-ST (Franche-Comt\'e Electronique, M\'ecanique, Thermique et Optique - Sciences et Technologies), CNRS (Centre National de la Recherche Scientifique), Supmicrotech-ENSMM (Ecole Nationale Supérieure de Mécanique et des Microtechniques), Université de Franche-Comté, 25000 Besan\c con France, \\(e-mail: giordano@femto-st.fr)}
\thanks{Manuscript created September 2023. ; This work is partially funded by the Agence Nationale de Recherche (ANR) Programme d'Inves\-tis\-se\-ment d'Avenir under the following grants:
   ANR-11-EQPX-0033-OSC-IMP (Oscillator IMP project)
    ANR-10-LABX-48-01 (FIRST-TF network), and
   ANR-17-EURE-00002 (EIPHI); 
   by the French RENATECH network and its FEMTO-ST technological facility,
  and by grants from the Région Bourgogne Franche Comté intended to support the above.}}


\maketitle

\begin{abstract}
Cylindrical WGM resonators machined in high-quality sapphire monocrystal cooled down to liquid helium temperature offer exceptionally-high Q-factors in the microwave frequency domain. Such a resonator constitutes the core of an ultra-stable oscillator featuring fractional frequency stability better than $1\times 10^{-15}$ at short integration times. As in any cylindrical resonant structure, the WGM resonator presents a two fold degeneracy. When a defect breaks the cylindrical symmetry of the resonator, the WGMs split and appear as doublets. In the high-quality sapphire resonator, the frequency separation of these twin modes varies from one mode order to another with a maximum value of a few tens of kHz. While the mode splitting for a given mode was considered until now unpredictable and intrinsic to each resonator since resulting a priori from randomly distributed defects. we show here, at the contrary, that the observed mode  splitting found on all the sapphire resonators whatever their origin mainly comes from a perfectly determined defect resulting from the manufacturing processes.
\end{abstract}


\section{Introduction}

\IEEEPARstart{W}hispering gallery modes (WGM)  low loss dielectric resonators cover a wide range of applications in the fields of microwaves, millimetre waves and optics. The strong confinement of the electromagnetic field inside the dielectric medium obtained with such excitation modes provides exceptional properties to the resonator such as: high quality factor, immunity to environment perturbations, enlarged dimensions ... In the microwave domain, these resonators serve as frequency reference for the demonstration of ultra-stable cryogenic oscillators \cite{cryogenics-2016,zhu-2018,locke08,vitusevich03}, and enable the accurate measurement of material complex permittivity \cite{krupka99-ieeemtt} or superconductors surface impedance \cite{cherpak-2003}. Millimeter-wave filters, oscillators or multiple-port power dividers/combiners have been developed using planar WGM dielectric resonators \cite{jiao-1987-whispering,cros-90}. While it is certainly in the field of photonics that we find today the most innovative applications of WGMs with a profusion of resonator shapes, sizes and operating wavelengths \cite{matsko-2006,ilchenko-2006}. \\ 

A wide variety of resonator shapes such as disks, rings, toroids or spheres, can support the propagation of whispering gallery modes. The rotational symmetry of these resonators induces a degeneration of the Helmhotz equation solutions yielding the two-fold degeneracy of the WGMs. Any defect that breaks the resonator symmetry lifts this degeneracy and a resonance line splitting is generally observed  \cite{mtt05-degenerescence,li2013}. The defect can be a geometrical imperfection, a particle sticked on the resonator surface or an inhomogeneity inside the dielectric bulk \cite{filipov1995,yi2011}. For many applications, this phenomenon is detrimental. In the case of a microwave WGM resonator oscillator, frequency jumps can occur between the two degenerate modes, and thus compromise the  oscillator stability \cite{dick94,mgwl00,ivanov2021}. At the opposite, the degeneracy and line splitting phenomenon can be exploited in photonics WGM resonators for the sensing of nanoparticles like contaminants or virus \cite{kippenberg2010,jiang2020}. In any case, these respective applications would benefit from improved control of the mode splitting and better understanding of its origin.\\

So far, it is generally accepted that the mode splitting observed in the sapphire microwave WGM resonator arises from randomly located defects that affect the resonator geometry or its homogeneity. In this paper, we refute this common though idea and show that the mode splitting originates from a perfectly determined geometrical defect. The latter is found to be observed on all the sapphire resonators, whatever their origin. This defect follows the 6-fold symmetry of the sapphire crystal and is linked to the different crystallographic planes, which react differently when machining the sapphire. \\

As a secondary benefit of our study, we bring here a modest answer to the question \it Can One Hear the Shape of a Drum? \rm \  posed by Mr. Kac in 1966 in a famous article \cite{kac-1966}. This question took mathematicians about thirty years to give in the general case a negative answer to this question \cite{gordon-1992}. It is also noteworthy that the first experimental verification of the isospectrality of two different geometries discovered by mathematicians was carried out using flat microwave cavities as a drum \cite{sridhar-1994}. In this article, we will obviously not address this complex mathematical problem. We show however how the phenomenon of "mode-splitting" observed in a microwave whispering gallery mode sapphire resonator enables to deduce its deviation from the ideal cylindrical form.\\ 

\section{The  Cryogenic Sapphire Resonator}
The mono-crystal of Al$_2$O$_3$ is the material presenting the lowest dielectric losses in the microwave frequency range. With a  loss tangent ($tg \delta$) of about $5 \times10^{-6}$ at  $300$~K and lower than $10^{-9}$ at liquid helium temperature \cite{braginsky87}, it is an ideal dielectric to build a high Q-factor microwave resonator in X-band. However, its relative permittivity is quite low, i.e. $\epsilon_r \approx 10$. 
Modes of high azimuthal order, named whispering gallery modes, are thus used to benefit fully from the sapphire's low dielectric loss tangent. A standard design consists of a large sapphire disk, typically 30 to 50 mm diameter (for X-band operation), enclosed in a metallic cylindrical cavity. For a whispering-gallery mode, total reflection occurs at the curved air–dielectric interface, greatly limiting the power dissipation in the metallic enclosure walls. The resonator unloaded Q-factor $Q_0$ is then only limited by the sapphire dielectric losses:
\begin{equation}
Q_0 \approx \dfrac{1}{tg \delta}
\end{equation}

Since 2010, we have built and validated at the FEMTO-ST Institute a dozen of ultra-stable Cryogenic Sapphire Oscillators (CSO) based on this technology. Our most advanced design is currently offered as a commercial instrument under the codename: ULISS-2G \cite{im-2023, apl-2023, www.uliss}. Describing the (in)stability as the Allan deviation (ADEV) $\sigma_y(\tau)$ of the fractional frequency $y$ as a function of the measurement time $\tau$ , ULISS-2G features $\sigma_y(\tau)< 3\times10^{-15}$ for $1~\text{s} \leq \tau \leq 10^4$~s and is limited by a drift of $\approx 10^{-14}$ at one day. It can run unattended for years of continuous operation, with only simple maintenance every 2nd year. \\

\begin{figure}[h]
\centering
\includegraphics[width=0.5\columnwidth]{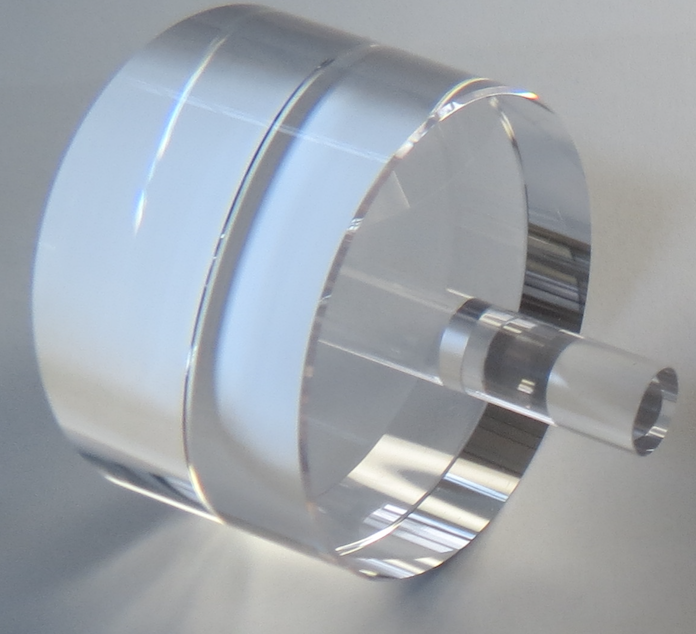}
\caption{\it \small The ULISS-2G sapphire resonator.}
\label{fig:fig1}
\end{figure}

All the resonators we have integrated into our CSOs are manufactured according to the same design shown in the figure \ref{fig:fig1} \cite{mtt-2015}. 

Made in a high purity sapphire mono-crystal, it is 54-mm-diameter and 30-mm-height with the cylindrical axis parallel to the Al$_2$O$_3$ crystal $c$-axis within 1$^\circ$. The $10$~mm diameter spindle machined from the bulk is used to attach the resonator to the bottom flange of a gold plated standard Oxygen-free high-thermal-conductivity (OFHC) copper cavity. The resonator operates on the quasi-transverse magnetic whispering-gallery mode
WGH$_{15,0,0}$ resonating at $\nu_0=9.99$~GHz \cite{rsi10-elisa}. Two diametrically opposed small magnetic loops constitute the input and output resonator ports. The radial position of these probes can be adjusted to tune the coupling coefficient at each resonator port. The cavity is thermally linked to the $2^{\text{nd}}$-stage of a Pulse-Tube cryocooler and stabilized near $6$~K. \\

\section{The mode degeneracy}

\begin{figure}[h]
\centering
\includegraphics[width=0.75\columnwidth]{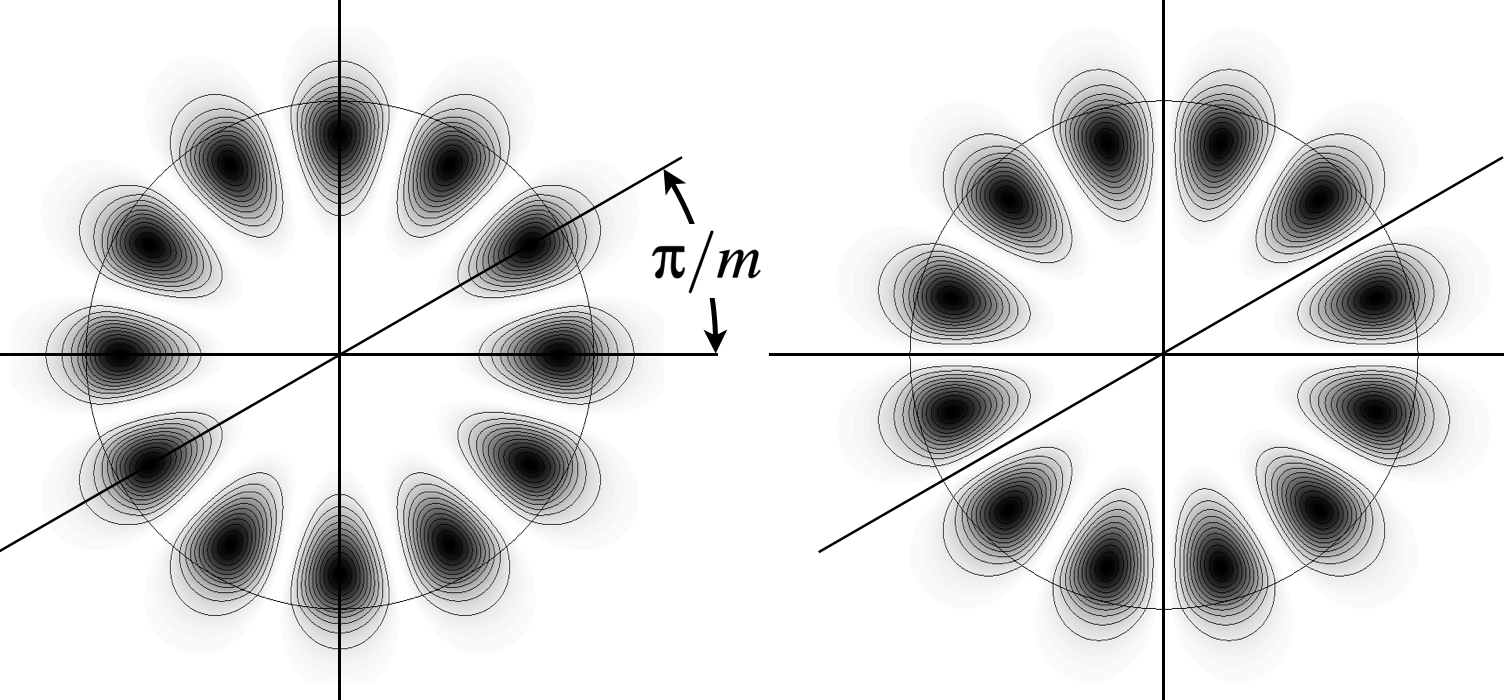}
\caption{\it \small Magnetic field in the resonator equatorial plane for the twin modes WGH$_{6,0,0}$.}
\label{fig:fig2}
\end{figure}

Although whispering-gallery (WG) modes are strictly hybrid modes, they can be classfied in quasi-TM (WGH) and quasi-TE (WGE) mode families. A WGH mode is characterized by the electric field mainly directed in the axial direction while its magnetic field is essentially transverse. The inverse situation occurs for a WGE mode. Whispering gallery modes can be further characterized by three integers, i.e. $m$,$n$ and $l$, representing the variation of the electromagnetic field components in the azimuthal ($\phi$), radial ($r$) and axial ($z$) directions of the cylindrical coordinate frame. In our developments, we only concentrate on the WGH$_{m,0,0}$ modes presenting the highest Q-factor. 
For these modes, the axial electric field inside the sapphire can be written as \cite{ivanov93}: 

\begin{equation}
E_z(r,\phi,z) = E_0 {J_m}(kr)\cos(\beta z) \left\{ \begin{array}{r}
        \!\!\!\!   \cos(m\phi) \\ 
         \!\!\!\! \sin(m\phi)
                \end{array}\right.\\
                \label{equ:Ez}
\end{equation}

where $J_m$ is the Bessel function of the first kind of order $m$, $\beta$  the axial propagation constant and $k$ the guided wave number. The brace indicates that the resonator can support two orthogonal "polarizations", both equivalent solutions of the Helmolzt equation.
Figure \ref{fig:fig2} shows the map of the transverse magnetic field in the resonator equatorial plane for the twin modes WGH$_{6,0,0}$.

In a perfect sapphire cylinder, the twin modes probe exactly the same medium, and thus resonate at the same frequency.
The position of the nodes of the stationary wave pattern is set by the coupling structure. It has already been shown how any defect affecting the resonator cylindrical symmetry lifts the twin modes degeneracy \cite{mtt05-degenerescence}. The frequency splitting $\delta \nu_m$ between the twin modes depends on the azimuthal number because $m$ conditions the configuration of the electromagnetic field near the defect. It is worth noting that in presence of the defect, the azimuthal position of the stationary wave pattern is no longer set by the coupling structure, but stays locked to the defect. One of the twin mode has a node at the defect location, the second one a maximum.\\

\begin{figure}[h]
\centering
\includegraphics[width=\columnwidth]{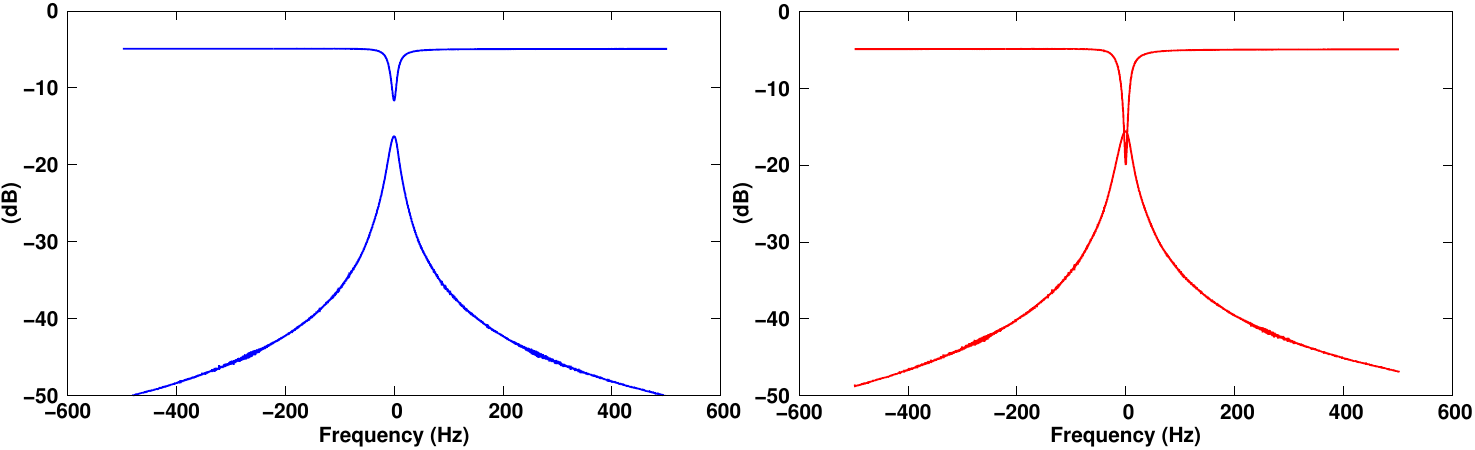}
\caption{\it \small Typical modulus of the reflexion and transmission coefficients $|S_{11}|$ and $|S_{21}|$ for the twin modes WGH$_{15,0,0}$ observed during the first cooling down to $6~$K. The insertion losses of the two modes are near the same, which can lead to frequency instability for the oscillator.}
\label{fig:fig3}
\end{figure}

In the CSO we exploit the mode WGH$_{15,0,0}$ at $\nu_{15}=~9.99$~GHz for which the twin resonances have always been observed. $\delta \nu_{15}$ is typically of the order of $10$~kHz ($10^{-6}$ in relative value).  
Figure \ref{fig:fig3} shows the two WGH$_{15,0,0}$ modes of a typical sapphire resonator cooled near $6$~K. Their frequency separation is $\delta \nu_{15}=8.6$~kHz ($ \frac{\delta \nu_{15}}{\nu_{15}}\approx 0.9\times 10^{-6}$). 

\begin{figure}[h]
\centering
\includegraphics[width=\columnwidth]{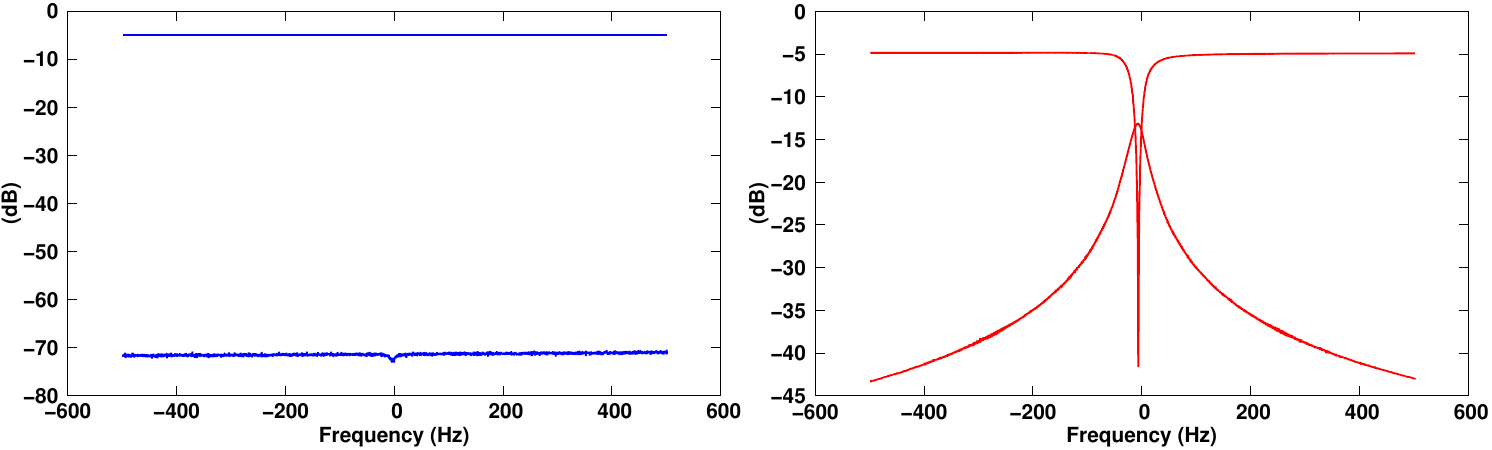}
\caption{\it \small Modulus of the reflexion and transmission coefficients $|S_{11}|$ and $|S_{21}|$ for the twin modes WGH$_{15,0,0}$ measured after a resonator rotation and a second cooling down to $6~$K. The sapphire rotation favoured the higher frequency mode, which is now the only mode that can oscillate.}
\label{fig:fig4}
\end{figure}

The presence of the twin modes is an issue for our application. Indeed, the oscillator will start on the mode having the lower insertion losses (higher coupling). The latter will be hardly predictable before cooling as the degeneracy lifting in not observable at room temperature. Under the worst conditions, both modes may have equivalent insertion losses. In this case the stability of the oscillator could be jeopardized by random frequency jumps.  This issue has been solved by rotating the sapphire resonator with respect to the coupling probe position such that the mode with the lowest frequency is suppressed, as demonstrated in the figure \ref{fig:fig4}.

Apart the WGH$_{15,0,0}$ mode, which serves as frequency reference for our oscillator, the experimental set-up enables the observation other WGH modes ranging from about 4 GHz to 14 GHz corresponding to $5\leq m\leq 24$. Outside this range, the limited bandwidth of the isolators placed at each resonator port and the poor coupling of the remaining modes make usually their detection impossible.   \\

\begin{figure}[h]
\centering
\includegraphics[width=\columnwidth]{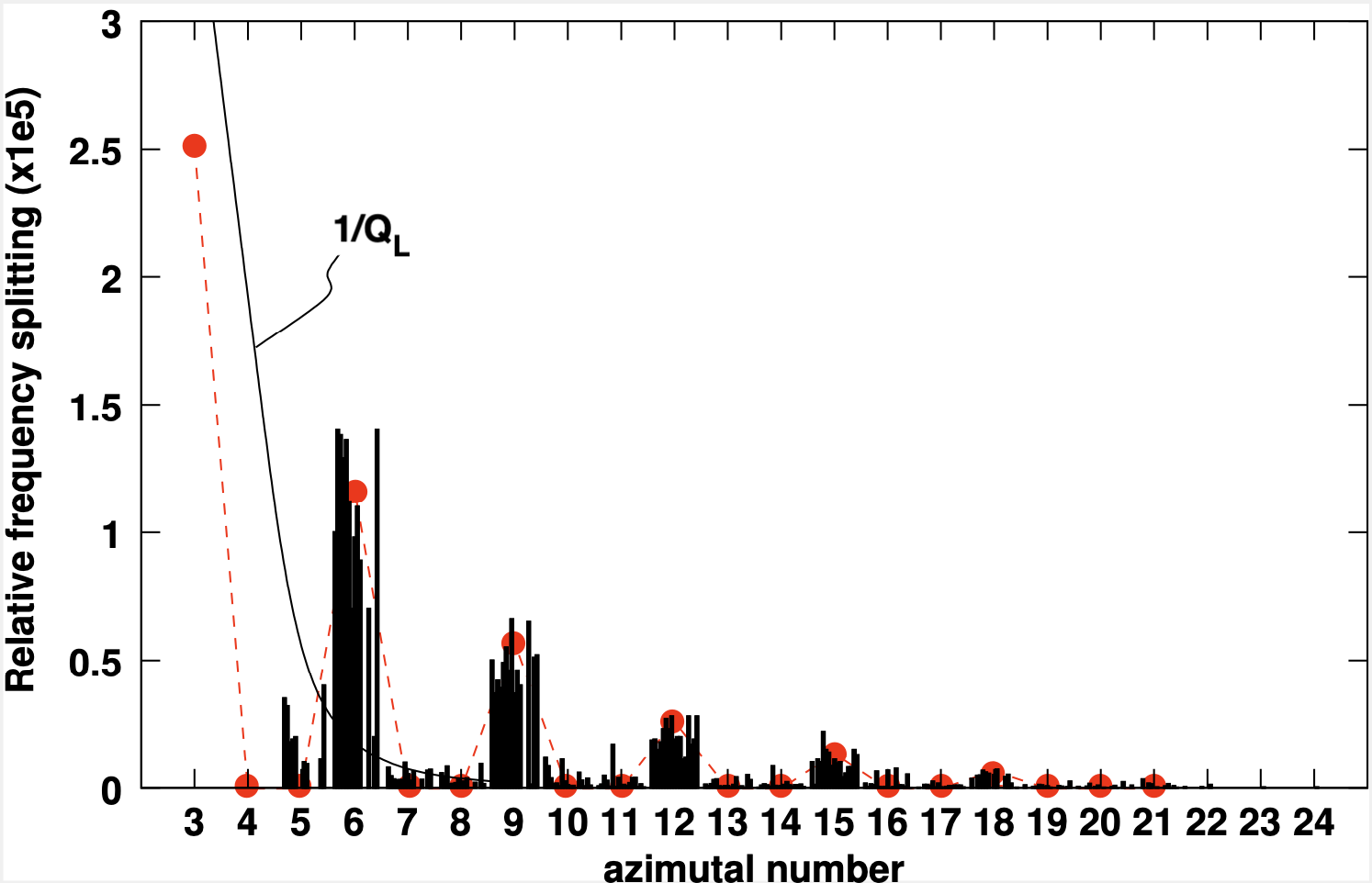}
\caption{\it \small Compilation of the relative mode splitting $\delta \nu_m/\nu_m$. For a given azimuthal number $m$, each resonator is represented by a black bar. The red bullets represent the calculation result obtained with $\Delta r =1.3~\mu$m (see section \ref{sec:model}).}
\label{fig:fig5}
\end{figure}

Until now we had not invested much on the origin of the mode splitting. It was generally accepted, that it results from defects affecting the cylindrical shape of the resonator or the morphology of the sapphire crystal. Since disturbances should be randomly distributed, the different mode splittings $\delta \nu_m$ are hardly predictable and should be specific to each resonator. To our opinion this claim must be abandoned as demonstrated in the figure \ref{fig:fig5}.\\

Figure \ref{fig:fig5} compiles the relatives values of $\delta \nu_m/\nu_m$ measured on more than fifteen resonators. We turn to manufacturers who manage all the processing operations: from crystal growth to final polishing. Among the tested resonators, some of them have been ordered more than ten years ago. These resonators come from three distinct manufacturers located in three different continents and using their own method of growth. Measurement results obtained with a resonator  of $50$~mm diameter and $30$~mm height and with another operating on the WGE modes are also included in the data.\\
 
\begin{figure}[h!]
\centering
\includegraphics[width=\columnwidth]{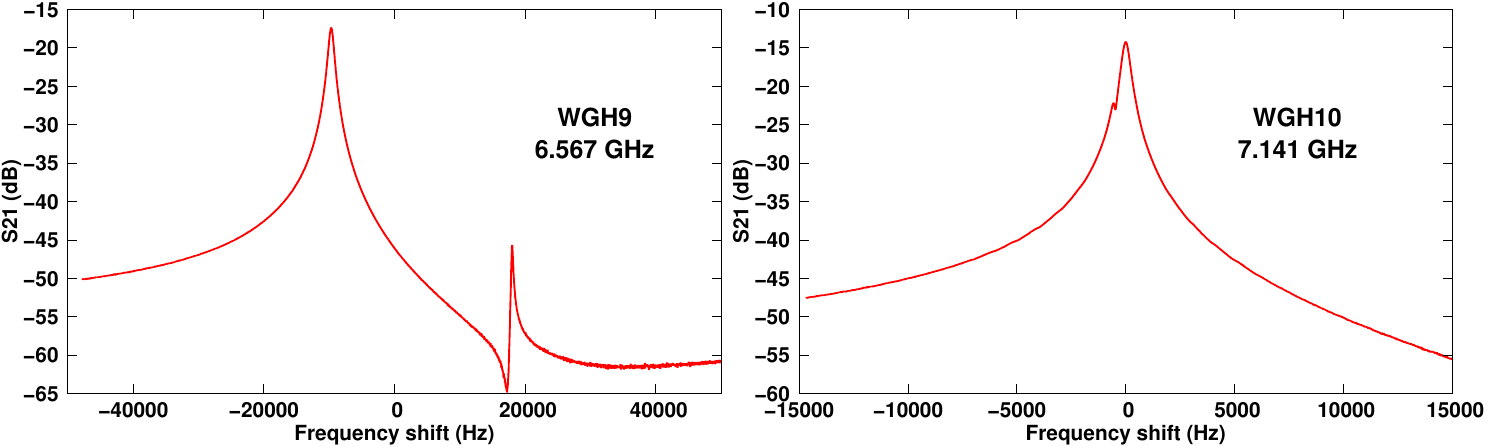}
\caption{\it \small Modulus of the transmission coefficient $|S_{21}|$ for the WGH$_{9,0,0}$ and WGH$_{10,0,0}$ modes at $6~$K.}
\label{fig:fig6}
\end{figure}

It appears obvious that the observed mode splitting distribution is not fully random. The mode splitting is notably higher for modes with $m$ multiple of 3 ($m=\text{mult}(3)$), regardless of the origin of the crystal, its dimensions or the excited mode family. In the observable frequency range, a maximum occurs for $m=6$ and then $\delta \nu_m$ decreases with $m$ following a curve parallel to $1/Q_L$, $Q_L$ being the loaded Q-factor. This last observation rather confirms the presence of a defect that affects the external shape of the resonator. Modes of high order being more confined in the dielectric are less affected by a defect on the resonator surface. For other modes with $m \neq \text{mult}(3)$, the mode splitting is generally not observed or stays of the order of the mode bandwidth. Figure \ref{fig:fig6} shows the resonator transmission coefficient $|S_{21}|$ for the two modes WGH$_{9,0,0}$ and WGH$_{10,0,0}$ measured at $6~$K. The mode splitting $\delta \nu_9 \approx 30~$ kHz is here clearly resolved, whereas for $m=10$ the  twin modes can be hardly distinguished.

\section{Deductive simplified model}
\label{sec:model}

 It is unlikely that the same defects are found on all these resonators of very different origin, unless these defects are related to an intrinsic property of the sapphire crystal.
Lu et al. showed for a photonics WG modes micro-cavity that an intentionally modulated resonator diameter induces a controlled mode splitting for a selected azimuthal mode number \cite{lu2020}.
The fact that modes with $m=\text{mult}(3)$ have the largest observed mode splitting suggests that there may be a periodic cylindrical defect that would mostly interact with this class of modes.\\

\begin{figure}[h!]
\centering
\includegraphics[width=0.45\columnwidth]{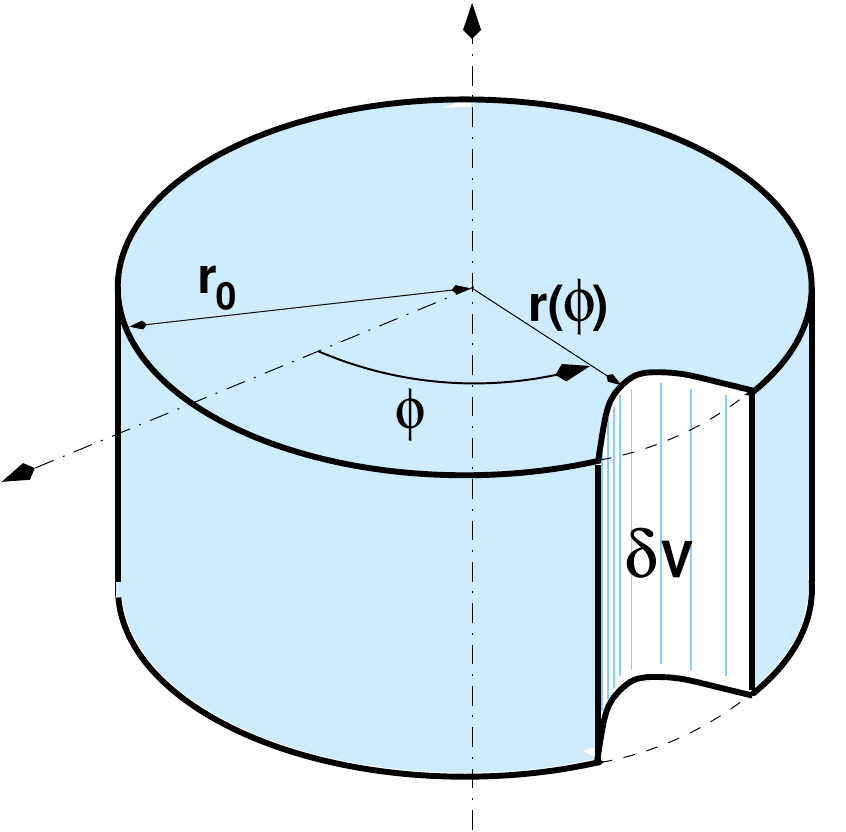}
\caption{\it \small Whispering Gallery Mode resonator with a small geometrical defect.}
\label{fig:fig7}
\end{figure}

Let's consider a resonator with a small geometrical defect as represented in the figure \ref{fig:fig7}. 

The resonance frequency shift resulting from this geometrical defect can be evaluated with the perturbation method \cite{argence1964}.
If the electric $\bf E$ and magnetic $\bf H $ fields in the unperturbed resonator are known, the relative frequency shift can be calculated by applying:
\begin{equation}
\dfrac{\Delta \nu}{\nu_{0}}=\epsilon_0 \dfrac{\iiint_{\delta V}  \bf{E}\rm \left( [\epsilon] - 1 \right  ) \bf{E^{*}}\rm  dv   } {\iiint_{V}  \left( \mu_{0}\bf{H}\rm ^{2}+\epsilon_{0} \bf{E}\rm [\epsilon]\bf{E^{*}}\rm \right  ) \rm dv}
\label{equ:perturb}
\end{equation}

where $\epsilon_0$ and $\mu_0$ are the vacuum permittivity and permeability respectively, $[\epsilon]$ the medium permittivity tensor, $V$ the entire resonator volume and $\delta V$ is the volume affected by the geometrical defect. $\delta V >0$ corresponds to a dielectric removal.  We have at our disposal a custom software based on the Mode Matching Method \cite{ivanov93}, which enables to calculate the frequency and the electromagnetic field components for any cylindrical WGM resonator.\\

The expression (\ref{equ:perturb}) can be simplified by considering that for a WGH mode, the electric field is principally axial: $\bf {E}\rm\approx E_z(r,\phi,z)\bf z$. Moreover, as the deformation is small with respect to the sapphire dimensions, $J_m(kr)$ can be considered as a constant over $\delta V$. Taking in account these approximations, it is can be shown from the expressions (\ref{equ:Ez}) and (\ref{equ:perturb}), that for a given azimuthal number $m$, the twin modes will be  differently  shifted and $\delta \nu_m$ is proportional to:
\begin{equation}
 I_m=  \int_0^{2\pi}\cos (2m\phi)d\phi  \int_{r(\phi)}^{r_0} r dr
\end{equation}

Let us consider a small modulation of the resonator radius such as: $r(\phi)=r_0+\Delta r \cos(p\phi)$, with $p$ being an integer and $\Delta r \ll r_0$. $I_m$ becomes:

\begin{equation}
I_m=2 r_0 \Delta r \int_0^{2\pi} \cos (2m\phi) \cos(p\phi) d\phi
\end{equation}

$I_m$ takes a non null value only if $p=2m$. As a first consequence, $p$ should be an even integer to induce a mode-splitting. Thus the radius modulation only affects the mode with $m=p/2$ , leaving the other WGH modes unperturbed.  
To obtain a distribution like those of the figure \ref{fig:fig5}, the radius modulation should be less \it{smooth}\rm. In other words, $r(\phi)$ should contain harmonics of the initial modulation:
\begin{equation}
r(\phi)= r_0+\Delta r \sum_{j\geq1} a_j \cos(j \times p\phi)
\end{equation}
In that condition, $I_m$ becomes the pondered summation of terms proportional to the Fourier coefficients $a_j$. The $j^{\text{th}}$ term takes a non null value for a specific azimuthal number:
\begin{equation}
m_j=j\dfrac{p}{2}
\end{equation}

Thus, for a given even integer $p$, the azimuthal numbers for which a mode splitting appears follow an arithmetic progression with a reason $\Delta m$:
\begin{equation}
\Delta m= m_j-m_{j-1}=\dfrac{p}{2}
\end{equation}

In our case, we have  $m=\text{mult}(3)$, thus $\Delta m=3$ and:
\begin{equation}
p=6
\end{equation}

From this simple observation, we deduce that the geometric defect must have a periodicity of $\pi/3$ in the azimuthal direction to induce a mode-splitting distribution compatible with the experimental observations.\\

The shape of the resonator is not yet known. We have now to determine the coefficients of the development of $r(\phi)$.
The non null terms in $I_m$ is proportional to the Fourier coefficients $a_j$ such as $j=m/3$. The measured $\delta \nu_m/\nu_m$ gives thus an estimation of the relative value of the Fourier coefficients $a_j$. 
 As previously mentioned, the mode splitting for $m=3$ can not be experimentally determined. Thus, we assume for the doublet WGH$_{3,0,0}$ a fractional mode splitting $\delta \nu_3/\nu_3=3\times 10^{-5}$ corresponding to the typical mode half bandwidth. For the other $m$ values, we take the average of the collected data. Setting $a_1=1$, we get for the Fourier coefficients up to $j=6$:\\
 
\begin{table}[h]
\centering
\caption{Fourier coefficients of the expected radius deformation $r(\phi)$}
\begin{tabular}{cccc}
$j$	&$m$		& $\delta \nu/\nu_m$	& Fourier Coef.\\
\hline 
1	&3		&$(3.0\times 10^{-5})$	&$a_1= 1.000 $\\
2	&6		&$1.0\times 10^{-5}$		&$a_2 = 0.333 $\\
3	&9		&$4.5\times 10^{-6}$		&$a_3 = 0.150 $\\
4	&12		&$2.0\times 10^{-6}$		&$a_4 = 0.067 $\\
5	&15		&$1.0\times 10^{-6}$		&$a_5 = 0.034 $\\
6	&18		&$5.0\times 10^{-7}$		&$a_6 = 0.017$ \\
\hline
\end{tabular}
\end{table}

\begin{figure}[h!]
\centering
\includegraphics[width=0.5\columnwidth]{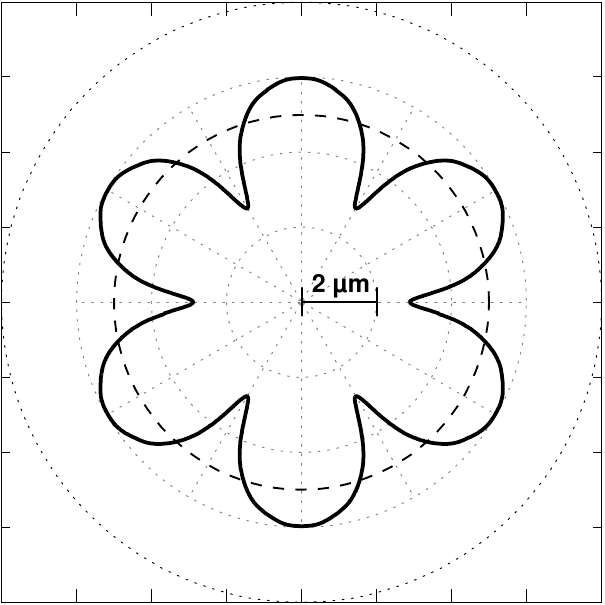}
\caption{\it \small Resonator profile deduced for the measurement of $\delta \nu_m/\nu_m$ assuming $\Delta r=1.3~\mu$m. The dashed line represents the nominal resonator diameter. }
\label{fig:fig8}
\end{figure}

The amplitude of the defect $\Delta r$ remains the only free parameter. We computed for different values of $\Delta r$ and for each azimuthal number $m$, the mode splitting using Equ.~(\ref{equ:perturb}). Here, the previous approximations are not used: we take into account of all electromagnetic fields components, their variation with respect to the coordinates and the anisotropy of the dielectric medium. In Fig.~\ref{fig:fig5} the red points represent $\delta \nu_m/\nu_m$  calculated with $\Delta r=1.3~\mu$m, which gives the best fit with the experimental observations. The corresponding resonator shape is represented in Fig.~\ref{fig:fig8} in polar coordinates.

The shape of the resonator deduced from our calculation is surprising at the first glance. The calculated resonator profile looks like a flower with 6 well marked petals. The peak-valley radius deviation is $\approx 3~\mu$m. We will see in the next section that the real profile of the resonator is actually similar to this one. We will also provide an explanation of this specific shape.

\section{Actual resonator profile}
\begin{figure}[h!]
\centering
\includegraphics[width=\columnwidth]{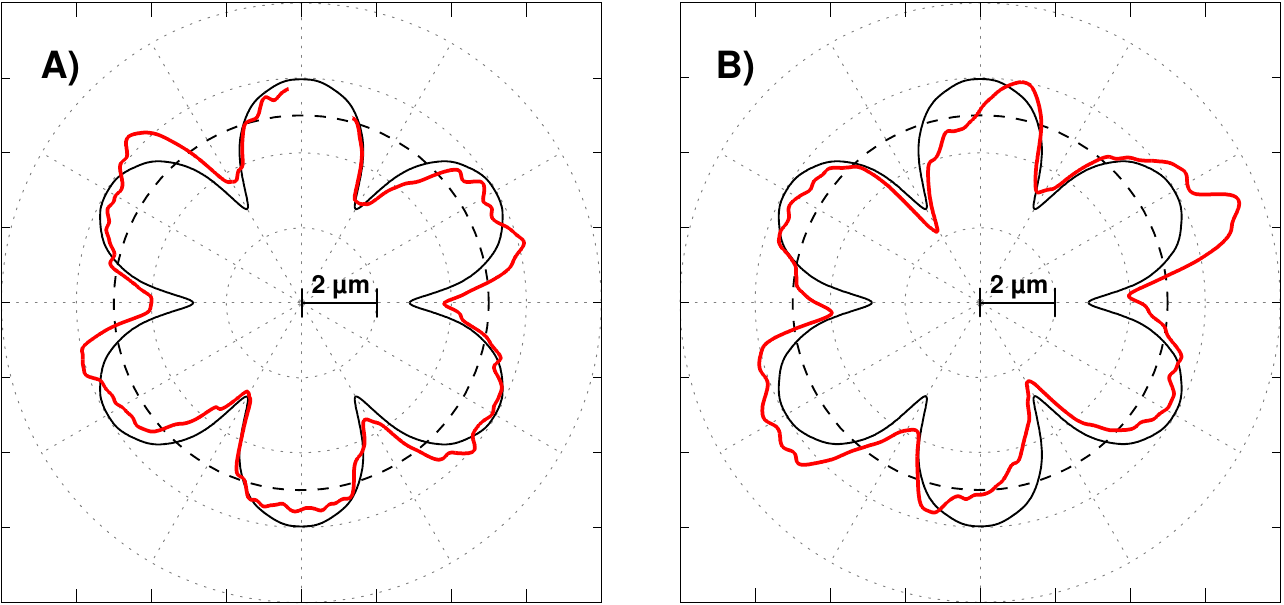}
\caption{\it \small Red plain line: Profile of two sapphire resonators measured with a resolution of $0.1~\mu$m. Black plain line: Profile deduced from frequency measurements. Dashed black line: Resonator nominal diameter.}
\label{fig:fig9}
\end{figure}

The profile of two sapphire resonators from different manufacturers have been measured using a metrological optical coordinate measuring machine $\mu$CMM Bruker Alicona. With this instrument, the resonator contour was measured in its equatorial plane with a resolution of $0.1~\mu$m. Fig.~\ref{fig:fig9} shows the two measured profiles compared to the modeled one (see Fig.~\ref{fig:fig8}).

It is remarkable that the calculated profile faithfully follows the real contour of the resonator.
This comparison fully validates the calculation presented in the previous section. The radius of the two resonators is modulated with a periodicity of $\pi/3$, and the peak-valley radius deviation is about $3~\mu$m, and this same defect is found on all resonators whatever their origin. It is therefore natural to think that this defect comes from an intrinsic property of  the sapphire crystal polishing.

\section{Proposed explanation}

\begin{figure}[h!]
\centering
\includegraphics[width=.8\columnwidth]{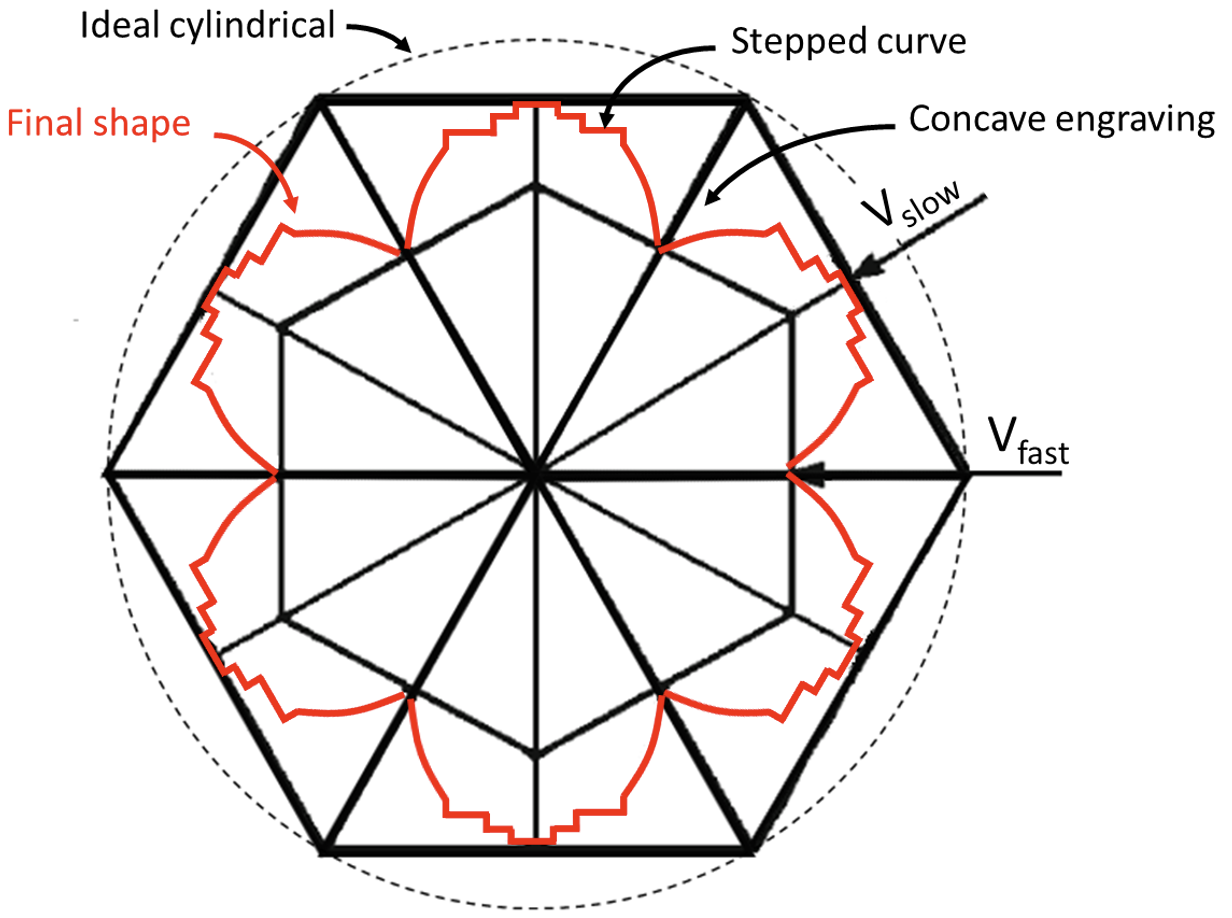}
\caption{\it \small Polar diagram representing the crystallographic planes, the etching rates and the expected final shape (figure adapted from \cite{dobrovinskaya2009sapphire}).}
\label{fig:fig10}
\end{figure}

In the following, a discussion on the intriguing contour of the sapphire resonator is presented from a manufacturing and material perspective. All $c$-oriented sapphire cylinder were prepared by chemical mechanical polishing (CMP) composed of particles and chemical agent with rotation polishing machine (whose exact production conditions may change from one manufacturer to another). However, the contour presents systematically large curved surfaces and concave engraving defects as shown in Fig.~\ref{fig:fig9} that are schematically redrawn in Fig.~\ref{fig:fig10}. Symmetry of order 6 indicates that one of the prismatic planes $a~(11\overline{2} 0)$ or $m~ (10\overline{1}0)$ planes (the first two densest planes perpendicular to the $c$-plane~(1000)) is favoured over to the other. Furthermore, looking carefully at Fig.~\ref{fig:fig9}, the curved surfaces show a stepped shape indicating that it corresponds to a crystallographic plane which tends to maintain its flat surface (low surface energy). This plane has the slower CMP etching rate (noted $V_{slow}$) than any other prismatic plane of sapphire. As for the second family of prismatic plane, located at $\pi/6$ to the first one, it presents a concave engraving. This family plane has the fastest polishing rate (noted $V_{fast}$). The engraving phenomena can be explained by the Rehbinder effect \cite{lee2020current}. The polishing chemical agent with the highest surface energy tends to accumulate the polishing particles to this face under the polishing pressure. The particles create microcracks after microcracks penetrating the crystal \cite{dobrovinskaya2009sapphire}. Figure \ref{fig:fig10}  schematizes the polar diagram of the crystal shape with two CMP etching rates. At this level of discussion, the determination of faster and slower CMP etching rate surfaces among the $a$- or $m$-planes has not been determined by diffraction X-rays in this study.
\cite{smith1996sapphire}\\

The literature presents several conflicting results on the CMP polishing rate of $a$- and $m$-planes \cite{zhu2004chemical,wang2020material}, as well as crystal morphology, surface energy, chemical dissolution and tribological/mechanical tests. Additionally, CMP polishing rate may likely change behavior due to surface atomic relaxation, crystal impurities, surfactant composition, water, wear rotation speed, or temperature \cite{dobrovinskaya2009sapphire}. Meanwhile, the surface energy is generally correlated to the friction but also to the hardness and etching rate \cite{dobrovinskaya2009sapphire}. As a matter of fact, the two prismatic planes may have a very close CMP polishing rate. Indeed, one can note that the final contour gives a very low crystal roughness of $3~\mu$m for a crystal diameter of $54$~mm which may depend on the polishing time. The relative tolerance of $5\times 10^{-6}$ in micro-machining is in the state of the art for anisotropic crystal. Finally, another argument in favor of the Rehbinder effect can be seen in figure \ref{fig:fig9}. Almost all stepped shape and engraving defects have an asymmetrical shape tilted in the same direction, which is likely the direction of polishing rotation.

\section{Summary}
In this paper we have solved the problem of the origin of the mode splitting effect in the cylindrical sapphire whispering gallery mode resonator. The same mode splitting repartition as a function of the azimuthal mode order is effectively observed in any sapphire resonator whatever their origin. Thus, we deduced that the mode splitting occurs due to a radius 6 fold modulation of the resonator contour resulting from the resonator polishing step. The observed mode splittings enable to calculate an approximate resonator profile, which is confirmed with high resolution resonator contour measurements. 

\section{Acknowledgements}
The authors would like to thank the  team of the MIPHySTO (MIcrofabrication, Hybridation de Syst\`emes Techniques et d'Outillage) platform who realized the contour measurements and Rodolphe Boudot for its review.

\end{document}